\documentclass[aps,pre,amsmath,amssymb,color,pdflatex,citeautoscript,superscriptaddress,notitlepage,twocolumn,nofootinbib,tighten,letterpaper]{revtex4-2}
\usepackage[english]{babel}
\usepackage[T1]{fontenc}
\usepackage{unixode}
\usepackage{graphicx}
\usepackage{changepage}
\usepackage{xcolor}
\usepackage{dcolumn}
\usepackage{comment}
\usepackage{amsmath,amssymb,bbold,bm}
\usepackage{hyperref}
\usepackage{amsfonts}
\usepackage{listings}
\usepackage{braket}
\usepackage{float,latexsym}
\usepackage{multirow}
\usepackage{tikz}
\usepackage{color}
\usepackage{xr}
\usepackage{siunitx}
\usepackage[normalem]{ulem}
\usepackage{notes2bib}
\usepackage{bibunits}
\usepackage{cleveref}
\usepackage{physics}
\usepackage{blindtext}
\usepackage{comment}
\usepackage[version=4]{mhchem}

\newcommand{\figureone}{
	\begin{figure}[t!]
		\centering
		\includegraphics[width=\columnwidth]{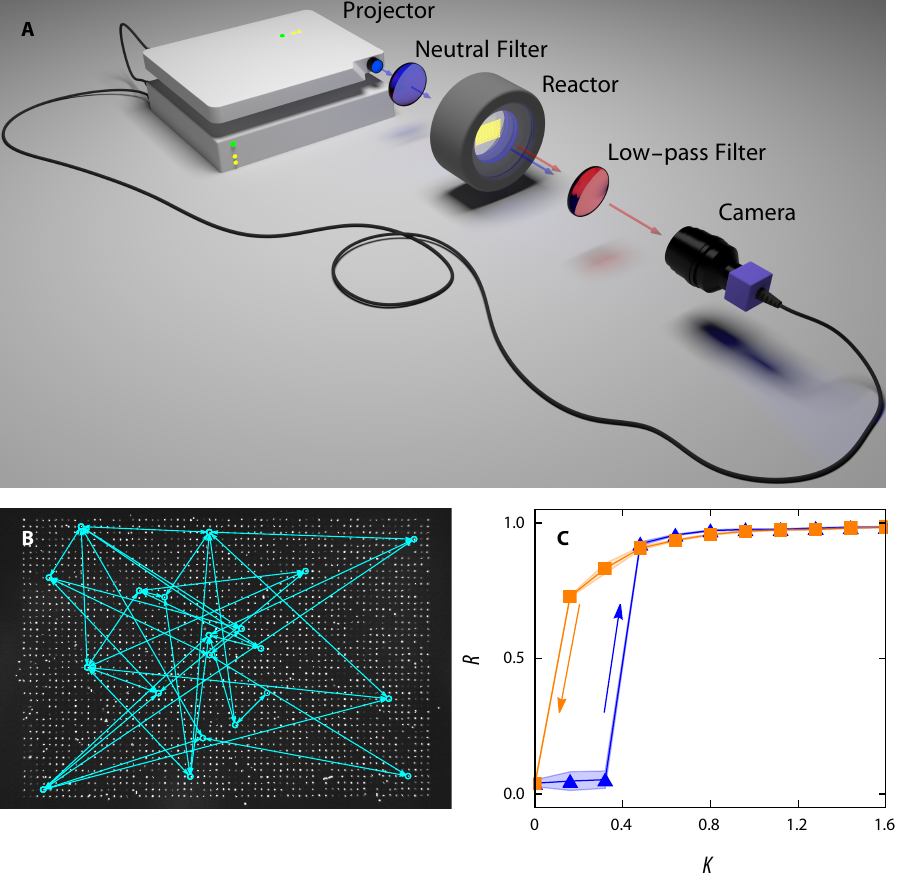}
		\caption{\textbf{Coupled photochemical oscillators.} (\textbf{A}) Experimental setup. A thermostatted open reactor, hosting 2600 chemical oscillators, is spectrophotometrically monitored in fluorescence light ($\lambda > \SI{550}{\nano\meter}$) with a CMOS camera. Recorded light intensities $f_i$ determine the photochemical feedback $I_i$, which is applied with a spatial light modulator. (\textbf{B}) Camera image of the fluorescing oscillator reservoir. The connectivity between oscillators is overlaid in blue. (\textbf{C}) Hysteresis loop of the Kuramoto order parameter $R$ in the case of $N = 1000$ all-to-all coupled oscillators with unimodally distributed natural frequencies (SD, $\sigma_\omega =\SI{0.012}{\radian\per\second}$).} 
	\end{figure}
}

\newcommand{\figuretwo}{
	\begin{figure*}[t!]
		\centering
		\includegraphics[width=0.75\textwidth]{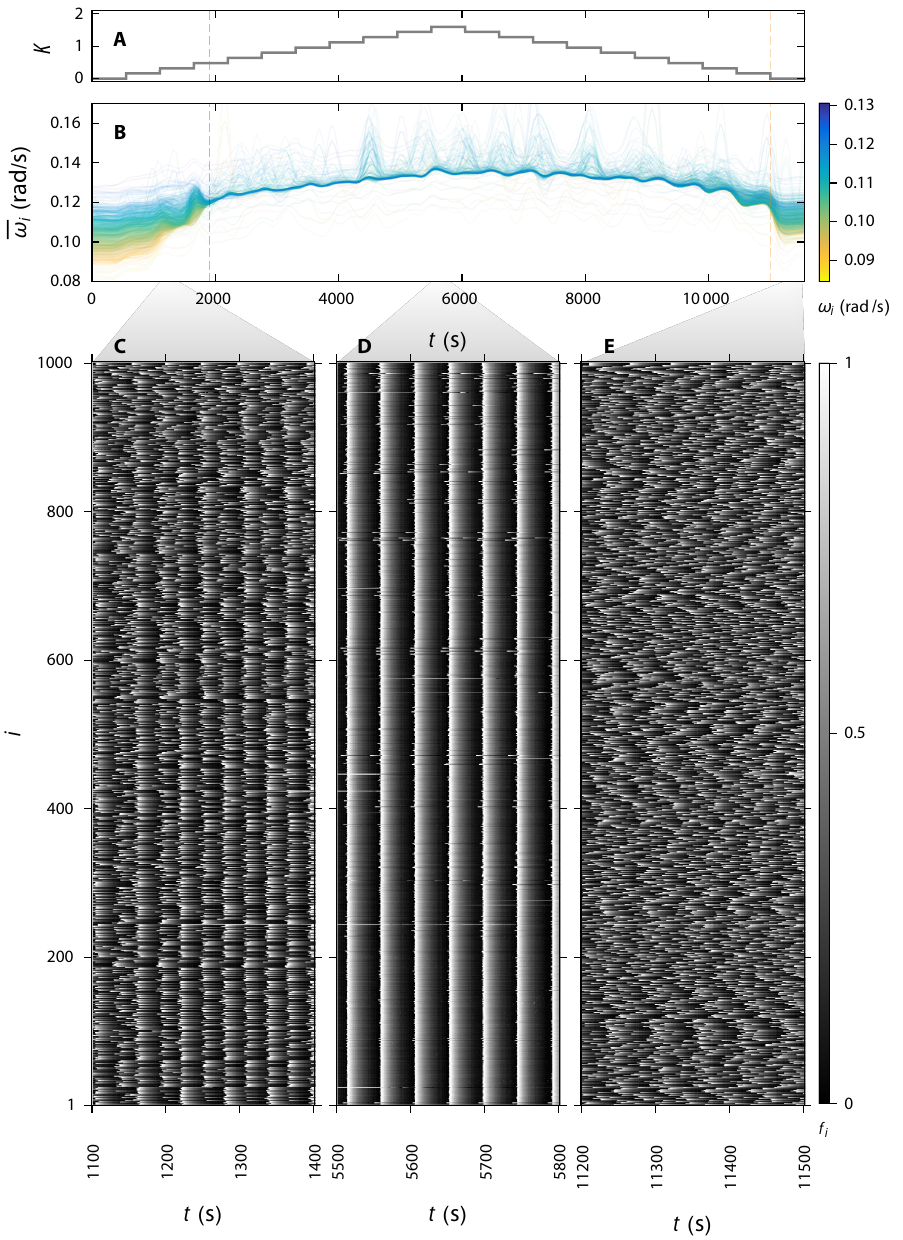}
		\caption{\textbf{Experimental observation of hysteresis in the case of $N = 1000$ all-to-all coupled oscillators with normally distributed natural frequencies.} (\textbf{A}) Time protocol for the coupling strength $K$. (\textbf{B}) Time evolution of the instantaneous frequencies ($\bar{\omega}_i$). The color of each line corresponds to the natural frequency of the nodes ($\omega_i$), respectively. The oscillators synchronize in-phase at $K_{\uparrow} = 0.4$ but transition back to incoherence at $K_{\downarrow} = 0.1 < K_{\uparrow}$ (see also Fig. 1C). (\textbf{C} to \textbf{E}) Fluorescence value ($f_i$) plots for clustering, synchronized, and incoherent states as observed during the cycle. The oscillators are indexed ($i$) in order of increasing natural frequencies. Synchronization from incoherent initial conditions (E) proceeds with the formation of antiphase clusters (C), which delay the onset of synchronization (D) leading to hysteretic behavior. The antiphase clustering state is characterized by a higher degree of frequency coherence for low-frequency oscillators.} 
	\end{figure*}
}

\newcommand{\figurethree}{
	\begin{figure*}[t!]
		\centering
		\includegraphics[width=0.75\textwidth]{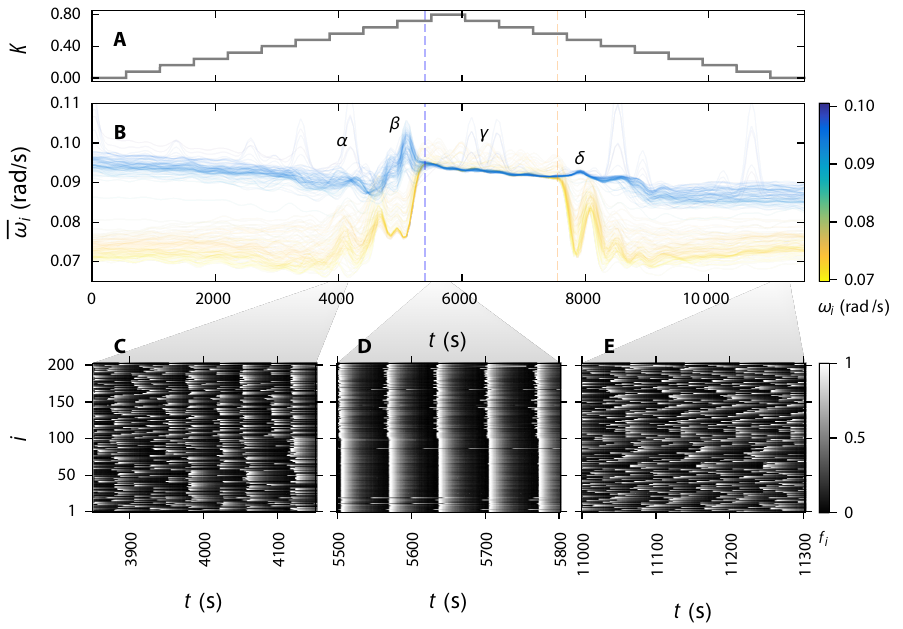}
		\caption{\textbf{Experimental observation of hysteresis for a bimodal frequency distribution.} (\textbf{A}) Time protocol for coupling strength $K$. (\textbf{B}) Time evolution of the instantaneous frequencies ($\bar{\omega}_i$) of $N = 200$ oscillators. The color of each line encodes the natural frequency of the nodes ($\omega_i$), respectively. The oscillators synchronize in-phase at $K_{\uparrow} = 0.72$ upon increasing $K$ but transition back to incoherence at $K_{\downarrow} = 0.56 < K_{\uparrow}$ upon decreasing $K$. (\textbf{C} to \textbf{E}) Individual fluorescence values for clustering (C), synchronized (D), and incoherent states (E) during the experiment. The oscillators are indexed ($i$) in order of increasing natural frequencies. Synchronization from incoherent initial conditions proceeds with the formation of approximately antiphase clusters, which delay the onset of synchronization leading to hysteresis.} 
	\end{figure*}
}

\newcommand{\figurefour}{
	\begin{figure}[t!]
		\centering
		\includegraphics[width=\columnwidth]{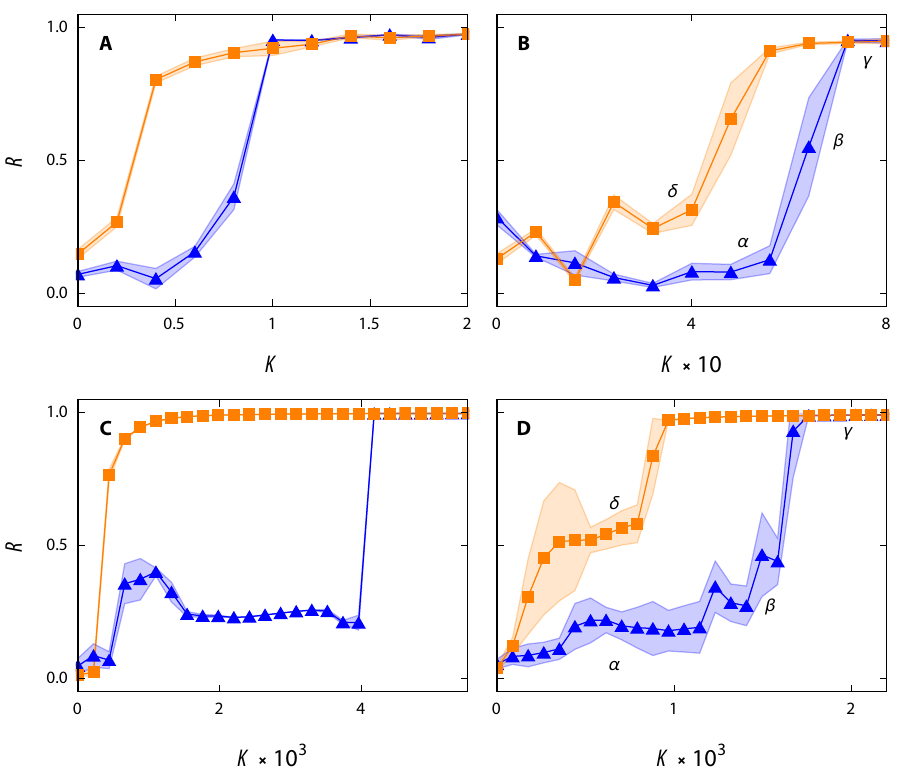}
		\caption{\textbf{First-order synchronization transitions in all-to-all coupled oscillator populations.} Hysteresis curves for the order parameter ($R$) with increasing (blue) or decreasing (orange) coupling strength in chemical experiments (\textbf{A} and \textbf{B}) and numerical simulations (\textbf{C} and \textbf{D}). We consider $N = 200$ globally coupled oscillators with unimodal (A and C) or bimodal (B and D) natural frequency distributions. Because of finite-size effects, $R \sim \mathcal{O}\left( 1/\sqrt{N} \right)$ in the unsynchronized phase (low $K$).} 
	\end{figure}
}

\newcommand{\figurefive}{
	\begin{figure*}[t!]
		\centering
		\includegraphics[width=0.75\textwidth]{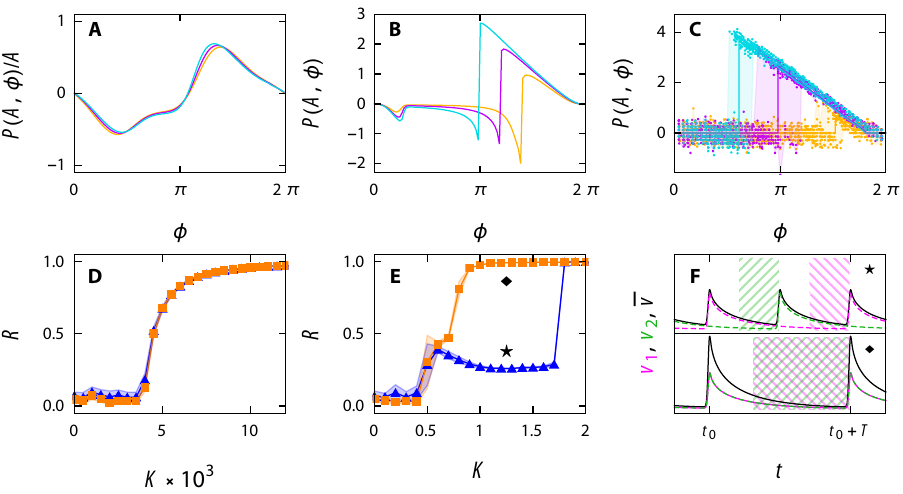}
		\caption{\textbf{Relaxation character determines the order of the synchronization transition.} (\textbf{A} and \textbf{B}) PRCs of the FHN model for time scale separation parameter $\epsilon = 0.3$ (A) with perturbation strengths of $0.4$, $0.6$, and $0.8$ (light-orange, purple, and light blue, respectively) and $\epsilon = 10$ (B) with perturbation strengths of $0.5$, $1.0$, and $1.5$ (light orange, purple, and light blue, respectively). The phase response curves are normalized by the respective perturbation strengths in (A). (\textbf{C}) Phase response curves determined from chemical experiments (dots) together with functions fitted with model (3) for perturbing light intensities of $0.01$ (light orange), $0.06$ (purple), and $\SI{0.25}{\milli\watt\per\centi\meter^2}$ (light blue). (\textbf{D} and \textbf{E}) Order parameter curves in the case of $N = 200$ globally coupled FHN oscillators with unimodally distributed natural frequencies for $\epsilon = 0.3$ (D) and $\epsilon = 10$ (E). (\textbf{F}) Schematic representation of the mechanism of first-order synchronization via antiphase cluster formation: time evolution of two oscillators from different subpopulations ($v_1$, $v_2$) and their excitable intervals (hatched regions) together with the mean amplitude ($\bar{v}$) in antiphase ($\star$) and in-phase synchronized ($\blacklozenge$) states.} 
	\end{figure*}
}

\begin{document}
\title{First-order synchronization transition in a large population of strongly coupled relaxation oscillators}

\author{Dumitru C\u{a}lug\u{a}ru}
\affiliation{Cavendish Laboratory, University of Cambridge, J.J. Thomson Avenue, Cambridge CB3 0HE, UK}
\affiliation{Department of Physics, Princeton University, Princeton, NJ 08544, USA}

\author{Jan Frederik Totz}
\affiliation{Institute of Theoretical Physics, Technical University Berlin, EW 7-1, Hardenbergstr. 36, 10623 Berlin, Germany}
\affiliation{Department of Mechanical Engineering, Massachusetts Institute of Technology, Cambridge, MA 02139, USA}
\affiliation{Department of Mathematics, Massachusetts Institute of Technology, Cambridge, MA 02139, USA}

\author{Erik A. Martens}
\affiliation{Department of Applied Mathematics and Computer Science, Technical University of Denmark, Richard Petersens Plads, 2800 Kgs. Lyngby, Denmark}

\author{Harald Engel}
\affiliation{Institute of Theoretical Physics, Technical University Berlin, EW 7-1, Hardenbergstr. 36, 10623 Berlin, Germany}
\date{\today}

\begin{abstract}
Onset and loss of synchronization in coupled oscillators are of fundamental importance in understanding emergent behavior in natural and man-made systems, which range from neural networks to power grids. We report on experiments with hundreds of strongly coupled photochemical relaxation oscillators that exhibit a discontinuous synchronization transition with hysteresis, as opposed to the paradigmatic continuous transition expected from the widely used weak coupling theory. The resulting first-order transition is robust with respect to changes in network connectivity and natural frequency distribution. This allows us to identify the relaxation character of the oscillators as the essential parameter that determines the nature of the synchronization transition. We further support this hypothesis by revealing the mechanism of the transition, which cannot be accounted for by standard phase reduction techniques.
\end{abstract}

\maketitle

\section{Introduction}
Since C. Huygens’s discovery of synchronization in coupled pendulum clocks in 1665~\cite{PIK01}, emergent synchronization of oscillating units has been observed in a myriad of natural systems, including firing neurons~\cite{EGG15,HAM07}, contracting cardiomyocytes~\cite{NIT16}, quorum-sensing bacteria~\cite{DAN10}, beating cilia~\cite{LIU18a}, rainfall extremes~\cite{BOE19}, and neutrino oscillations~\cite{PAN98}. In addition, synchronization underpins the dynamics of a variety of technological systems such as power grids~\cite{BUL10}, bridge instabilities~\cite{STR05}, traffic patterns~\cite{KER97}, and lasers~\cite{HIL20,SOR16}. The process of synchronization can be interpreted from a statistical mechanics perspective as a nonequilibrium phase transition, where a synchronized state emerges from an incoherent one as the coupling between the oscillators is increased. The seminal works of Winfree and Kuramoto~\cite{WIN67,KUR84} and subsequent experiments~\cite{KIS02} have shown that in populations of weakly coupled phase oscillators with a unimodal natural frequency distribution, such a transition proceeds continuously and reversibly in a second-order fashion. Under special conditions, the synchronization transition can also be of first-order type~\cite{PAZ05,MAR09,CHA19,TAN97,LEY12}. In this case, synchronization abruptly ensues at a critical coupling strength $K_{\uparrow}$ but disappears below a coupling strength $K_{\downarrow} < K_{\uparrow}$, resulting in hysteresis~\cite{KUM15,BOC16}. Such a discontinuous transition has been hypothesized to play a role in the onset of anesthesia-induced unconsciousness~\cite{KIM16}, epileptic seizures~\cite{YAF15}, acoustical signal transduction in the cochlea~\cite{WAN16}, hypersensitivity in chronic pain~\cite{LEE18}, and memory processes~\cite{FEL11}. The simplifying assumptions underlying the Kuramoto phase oscillator model, however, render it unsuitable for understanding systems where strongly coupled relaxation oscillators prevail. To experimentally investigate the onset of synchronization in large ensembles with $N = 200$ and $N = 1000$ relaxation oscillators, we use the well-studied Belousov-Zhabotinsky (BZ) chemical reaction~\cite{EPS16}. Despite intrinsic differences in the underlying microscopic mechanisms with biological neurons, this chemical reaction shows qualitatively identical emergent behavior: spiked slow-fast oscillations in the time traces of the concentrations and phase-dependent excitable response to external perturbations~\cite{TOT18,PRI03,NAB13,IZH07,ISO17}.

\section{Results and Discussion}
\subsection{Chemical micro-oscillators}
\figureone
We synthesized individual oscillatory units from ion-exchange resin beads saturated with a photosensitive BZ reaction catalyst, ruthenium(II)-tris(2,2'-bipyridine-dimethylene)-chloride~\cite{TOT18,TAY09,TAY08,TIN12}. The particles were then immobilized under a hydrogel layer on an acrylic plate and immersed in a catalyst-free reaction solution. This resulted in a reservoir of uncoupled chemical micro-oscillators. The oscillators can be coupled together photochemically with the experimental setup presented in Fig. 1A. During an oscillation cycle, the catalyst varies periodically between its fluorescing and nonfluorescing oxidation states. The phase of each oscillator can thus be monitored by optically measuring the oxidized catalyst concentration via its fluorescence intensity $f_i$ with a complementary metal-oxide semiconductor (CMOS) camera. On the basis of these intensities, an individual photochemical feedback
\begin{equation}
	I_i(t)=I_0 + K \sum_{j=1}^{N} A_{ij}\left[f_j(t) - f_i(t) \right]
\end{equation}
is calculated and projected on each photosensitive micro-oscillator with a spatial light modulator. The natural frequencies $\omega_i$ of all oscillators are measured at the start of each experiment under a uniform background light of intensity $I_0$. Subsequently, a suitable subset of the oscillators is selected to obtain a desired frequency distribution. Different network connectivities can be implemented by choosing an appropriate adjacency matrix $A_{ij}$. Figure 1B depicts a typical camera image of the fluorescing bead reservoir during an experiment together with a chosen network graph superimposed to highlight the connected oscillators. During each experiment, the coupling strength $K$ is cycled from low to high values and back. We monitor the synchronization level using the Kuramoto order parameter~\cite{KUR84}
\begin{equation}
	R = \left\langle \abs{\sum_{j=1}^{N} e^{\mathrm{i} \phi_j(t)}} \right\rangle_t
\end{equation}
where $\left\langle \dots \right\rangle_t$ denotes the time average and $\phi_j (t)$ represents the phase of the $j$-th oscillator, which is calculated by linear interpolation between consecutive firing events. The order parameter $R$ ranges from $0$, when the phases are incoherent, to $1$, where all phases align perfectly. The onset of synchronization can be captured from the dependence of the order parameter on the coupling strength.

\subsection{Role of natural frequency distribution}
\figuretwo
An experimentally recorded order parameter curve is shown in Fig. 1C for the case of $N = 1000$ globally coupled oscillators with a unimodal distribution of natural frequencies. Upon increasing $K$, there is an abrupt transition to a highly synchronized state at a critical value of the coupling strength, $K_{\uparrow} = 0.4$. Once this phase is formed, it remains stable until the coupling strength $K$ is decreased below $K_{\uparrow}$ to a smaller value, $K_{\downarrow} = 0.1$. This hysteresis cycle is characteristic of a first-order phase transition. The evolution of the instantaneous frequencies for each oscillator (Fig. 2) confirms the hysteretic nature of the transition: For a time-reversal symmetric protocol of the coupling strength $K$ (Fig. 2A), the oscillators' frequencies evolve asymmetrically with respect to time reversal (Fig. 2B). Moreover, the fluorescence time plots (Fig. 2, C to E) indicate that the emergence of an in-phase synchronized state is preceded by the formation of antiphase clusters. In addition, the frequency of the oscillators in the in-phase synchronized state is approximately given by the frequency of the fastest unperturbed oscillators. This contrasts with the Kuramoto model, which predicts that the oscillators phase-lock at a frequency equal to the average natural frequency of the entire population. 

\figurethree
We investigated the robustness of the first-order synchronization transition with regard to the frequency distribution in an all-to-all coupled network of BZ oscillators with frequencies drawn from a bimodal distribution~\cite{MAR09,MIK04}. Incidentally, this reveals the hierarchy of the emergent synchronization dynamics (Fig. 3). While cycling the coupling strength $K$ up and down (Fig. 3A), the evolution of the frequencies is asymmetric in time, which again indicates hysteretic behavior (Fig. 3B). At the beginning of the experiment, the coupling strength is small, and the oscillators are desynchronized and incoherent. Upon a slight increase of the coupling strength, the fluorescence time plot (Fig. 3C) shows the presence of approximately antiphase clusters ($\alpha$) in the subpopulations associated with fast and slow intrinsic frequencies. Oscillator heterogeneity induces intercluster switching, so there is no perfect frequency synchronization~\cite{TAY08}. Before the onset of global synchronization, the low-frequency subpopulation achieves in-phase synchronization, while the high-frequency group remains approximately antiphase but displays an average increase in instantaneous frequency ($\beta$). Once the synchronized state is established at $t \approx \SI{5300}{\second}$ ($K_{\uparrow} = 0.72$, blue dashed line), the population oscillates with the natural frequency of the fastest oscillators ($\gamma$). This regime is characterized by almost perfect phase alignment, with the fast oscillators entraining the entire population (Fig. 3D). The destabilization of the synchronized state at $t \approx \SI{7800}{\second}$ ($K_{\downarrow} = 0.56$, orange dashed line) is initiated by the loss of frequency coherence in the slower subpopulation ($\delta$). At the end of the experiment, we recover the fully incoherent state that is also observed in the beginning for very low ($K < 0.2$) coupling strengths (Fig. 3E).

\subsection{Synchronization mechanism for relaxation oscillators}
\figurefour
To gain more insight in the mechanism of the observed first-order synchronization transition, we performed numerical simulations with an established model of the BZ chemical kinetics~\cite{TOT18,ZHA93}. Figure 4 shows the comparison of the hysteretic order parameter curves between experiments and simulations in the case of globally coupled oscillators with unimodal and bimodal distributions. In both cases, the ascending branch of the hysteresis loop is characterized by a persistence of low order parameter values. The detailed inspection of the collective node dynamics in the case of unimodal (Fig. 2 and fig. S1) and bimodal (Fig. 3) frequency distributions reveals that the abrupt emergence of an in-phase synchronized state is preceded by the formation of antiphase clusters. The antiphase synchronized state effectively suppresses the onset of in-phase synchronization, resulting in hysteretic behavior. The critical coupling strength for in-phase synchronization corresponds to the point where the antiphase state vanishes.

Moreover, we found that the collective transition for many oscillators can be described by a reduced model of just two coupled identical oscillators. In this case, direct simulations reveal that, up to a certain critical coupling strength, both in-phase and antiphase states coexist (fig. S2). Beyond a critical point, the antiphase state becomes unstable. An estimate for the critical coupling strength, $K^* = 4.3 \times 10^{-3}$, agrees quantitatively in the case of a unimodal distribution of natural frequencies and qualitatively for a bimodal distribution, where the effects of frequency distribution are more pronounced. The peculiarities of the latter case (the sequence of $\alpha$, $\beta$, $\delta$, and $\gamma$ states) are also accurately reproduced by the simulations in both the order parameter curves and fluorescence intensity plots (fig. S3).

\subsection{Network topology}
To further demonstrate the robustness of the first-order synchronization transition for relaxation oscillators, we investigated the role of network connectivity. We chose two paradigmatic random networks, the Barabási-Albert and the Erdős-Rényi graphs, where the natural frequencies depended linearly on the corresponding node degrees~\cite{BOC16}. Our experiments and simulations with chemical relaxation oscillators show that there is a discontinuous first-order transition to in-phase synchronization, with hysteresis irrespective of the chosen network connectivity models (fig. S4). This suggests that the occurrence of abrupt synchronization in the case of relaxation oscillators depends only weakly on the underlying network topology. Moreover, a close inspection of the collective node dynamics again shows the presence of approximately antiphase clusters suppressing the onset of in-phase synchronization (figs. S5 and S6).

\subsection{Relaxation dynamics and time scale separation}
To validate our hypothesis on the role of the relaxation character determining the type of synchronization transition, we use the FitzHugh-Nagumo (FHN) model~\cite{IZH07}, a canonical model for relaxation oscillations in the context of neuronal excitability. Varying the time scale separation, parameter $\epsilon$ allows for tuning between harmonic and slow-fast relaxation oscillations (Fig. 5). Every oscillator can be characterized by its phase response curve (PRC), which encodes the resultant phase change $\Delta \phi$ due to a short perturbation applied at a phase $\phi$~\cite{IZH07}. We observe that the PRC evolves from a linear to a nonlinear dependence on the perturbation amplitude $A$: While for negligible time scale separation, the PRC is roughly sinusoidal (Fig. 5A), for strong time scale separation, it is a discontinuous function whose jump point $\phi^* (A)$ shifts to smaller phases with increasing perturbation strength $A$ (Fig. 5B), thus enlarging the excitable interval. A simple approximation for the PRC is
\begin{equation}
	P(\phi,A) = \begin{cases}
		0 & \phi < \phi^* (A) \\
		2\pi-\phi & \phi \geq \phi^* (A) \\
	\end{cases}
\end{equation}

\figurefive
This PRC encodes phase-dependent excitability: At early phases, an oscillator is refractory, as perturbations do not affect it. However, in the excitable window with phases above $\phi^* (A)$, perturbations immediately trigger a new spike~\cite{PRI03,NAB13}. We observe identical behavior for our chemical oscillators (Fig. 5C). Simulations of a globally coupled network of $N = 200$ FHN oscillators indicate the presence of a continuous transition with-out hysteresis for negligible time scale separation ($\epsilon = 0.3$) and a discontinuous transition with hysteresis for a pronounced time scale separation ($\epsilon = 10$) (Fig. 5, D and E). In agreement with the experiments, individual dynamics of oscillators in the bistable region reveal antiphase states and in-phase states during the up- and down-sweep, respectively. This behavior is further confirmed in a reduced two-oscillator model, which shows the presence of stable antiphase states only in the case of pronounced time scale separation (fig. S7). The mechanism underlying the hysteretic synchronization transition for relaxation oscillators is thus revealed to be deeply rooted in the nonlinear behavior of the PRC (Fig. 5F). In an initially incoherent population of coupled oscillators at low $K$, a firing event from an oscillator (the pacemaker) triggers firing events in $n_1$ other oscillators, whose phases are in the excitable interval. As a result, the mean field amplitude $K \bar{v}(t)$ contains a spike amplified by a factor $n_1$, which will, in turn, induce even more ($n_2 > n_1$) oscillators to fire, since the excitable interval is now even larger (see Fig. 5, B and C). Thus, synchronization for relaxation oscillators can be seen to be driven by pacemakers, which are abundant due to the random initial conditions. The situation at low $\bar{v}$, where the excitable window of oscillators is short, will generally lead to the emergence of synchronized subpopulations that are effectively uncoupled. The simplest example is the antiphase state that is formed by two subpopulations with opposing phases: A firing event occurring in one subpopulation may entrain oscillators with similar phases belonging to the same subpopulation; however, the event cannot illicit spiking of the oscillators belonging to the other subpopulation, since they are in the refractory interval and cannot be excited. The width of the excitable interval grows with increasing coupling strength $K$ until it covers most of the oscillation cycle ($\phi^* (A) < \pi$) at the transition point ($K_{\uparrow}$), as illustrated in the bottom panel of Fig. 5F. In this situation, oscillators residing in one subpopulation can trigger responses in oscillators belonging to the other subpopulation, thus giving rise to in-phase synchronization across the entire system. In the $K$-decreasing branch, the in-phase synchronized state can persist for $K << K_{\uparrow}$; the already synchronized oscillators require only a small excitable interval to maintain their synchronized state. Last, the loss of synchronization at $K_{\downarrow}$ happens almost instantaneously, with the exact transition point depending on the frequency distribution.

\section{Summary And Outlook}
Our theoretical and experimental analysis points to the fact that simple phase models, while analytically tractable, can fail to capture emergent phenomena in ensembles of strongly coupled relaxation oscillators correctly. In the systems studied here, the type of synchronization transition is much more sensitive to the relaxation character than the frequency distribution or network connectivity as in the case of the weakly coupled phase oscillator model. Because of the ubiquitous nature of these systems, we expect the presented mechanism to play an essential role in understanding further oscillatory systems, such as next-generation neuromorphic photonic devices~\cite{HAR19} and optogenetically addressable neural networks~\cite{ADA19}, as well as in medical therapies of Tinnitus and Parkinson’s disease based on neural desynchronization strategies~\cite{EGG15,HAM07}.

\section{Materials And Methods}
\subsection{Preparation of the chemical oscillators}
Cation-exchange resin beads ($75$ to $\SI{150}{\micro\meter}$ in diameter; DOWEX WX4 100-200) were sieved to obtain a narrow size distribution ($116$ to $\SI{112}{\micro\meter}$). One gram of sifted beads was placed in $\SI{5}{\milli\liter}$ of water, and under constant vortex mixing, $\SI{15}{\milli\liter}$ of ruthenium(II)-tris(2,2'-bipyridine-dimethylene)-chloride (\ce{Ru(dmpy)3Cl2}) catalyst solution ($\SI{1.66}{\micro\mole\per\liter}$) was added slowly over the course of $\SI{10}{\minute}$. Mixing was continued for 48 hours until a homogeneous (verified by color saturation measurements) catalyst loading of resin ($\SI{2.5e-5}{\mol\per\gram}$) was achieved in all beads. The catalyst-soaked beads were placed on a drilled acrylic plate with a grid of $64 \times 44$ cylindrical wells (diameter, $\SI{200}{\micro\meter}$; depth, $\SI{150}{\micro\meter}$; separation, $\SI{400}{\micro\meter}$) and then evenly distributed with a fine brush by applying a water surfactant solution ($0.5 \% $ Triton X-100). After 3 hours, the beads were sealed with liquid silica hydrogel that solidified more than $\SI{30}{\minute}$~\cite{TOT18}. The chemical oscillations are started when the acrylic plate is sub-merged in a Belousov-Zhabotinsky reaction solution ($\left[ \ce{H2SO4} \right] = \SI{0.77}{\mole\per\liter}$, $\left[ \ce{NaBrO3} \right] = \SI{0.51}{\mole\per\liter}$, $\left[ \ce{NaBr} \right] = \SI{0.08}{\mole\per\liter}$, $\left[ \text{malonic acid} \right] = \SI{0.16}{\mole\per\liter}$).
\subsection{Data analysis}
During an experimental run, the coupling strength $K$ is cycled from low to high values and back. Each value of $K$ is maintained for $\tau_{\mathrm{coup}} = \SI{550}{\second}$ at the end of which the Kuramoto order parameter $R$ is determined by averaging over the past interval $\tau_{\mathrm{av}} = \SI{550}{\second}$ according to Eq. 3. Note that $\tau_{\mathrm{coup}}$ is large enough to ensure that the oscillators reach the (de-)synchronized steady state at the end of the respective coupling stage. Because of the phase-resetting nature of the relaxation oscillators, the oscillators synchronized within one or two periods after $K_{\uparrow}$ is reached (characteristic time scale of $1 / \omega_i \sim \SI{100}{\second}$). In a similar manner, desynchronization happens abruptly with the order parameter relaxing to its steady-state value on a characteristic time scale, which is inversely proportional to the spread of natural frequency ($1/\sigma_{\omega} \sim \SI{150}{\second}$). The instantaneous frequency of each chemical oscillator is computed directly from its temporal phase, $\bar{\omega}_i(t)=\dot{\phi}_i(t)$. As we are interested in the time evolution of $\bar{\omega}_i$ on time scales comparable with $\tau_{\mathrm{coup}}$, we convolve $\bar{\omega}_i$ with a Gaussian of width $\sigma_t = \SI{15}{\second}$ in Figs. 2 and 3 and figs. S1, S5, and S6.
\subsection{Numerical simulations}
For simulating the chemical kinetics of the BZ reaction, we use the nondimensionalized Zhabotinsky-Buchholtz-Kiyatkin-Epstein (ZBKE) model, which is further modified to account for photochemical effects due to coupling with light~\cite{TOT18,ZHA93}. The state of the $i$-th node is given by two variables, $u_i$ and $v_i$, which are proportional to the concentrations of the \ce{HBrO2} and \ce{Ru(dmbpy)3^3+} reaction intermediates, respectively. Their time evolution is given by
\begin{widetext}
\begin{align}
	\epsilon_1 \dot{u}_i &= I_i+\left(\frac{\alpha q v_i}{\epsilon_3 + 1 - v_i} + \beta \right) \frac{\mu - u_i}{\mu + u_i} + \gamma \epsilon_2 w^2_{i,\mathrm{SS}} + \left(1 - v_i \right) w_{i,\mathrm{SS}} - u_i^2 - u_i, \\
	\dot{v}_i &=2 I_i+\left( 1 - \nu \right) w_{i,\mathrm{SS}} - \frac{\alpha q v_i}{\epsilon_3 + 1 - v_i},
\end{align}
\end{widetext}
where
\begin{equation}
	w_{i,\mathrm{SS}} = \frac{1}{4 \gamma \epsilon_2} \left( \sqrt{16 \gamma \epsilon_2 u_i + v_i^2 -2 v_i + 1} + v_i - 1 \right) 
\end{equation}
represents the steady-state concentration of \ce{HBrO2+} and
\begin{equation}
	I_i = I_0 + K \sum_{j=1}^{N} L_{ij} \left[ v_j + \xi_{j} (t) \right].
\end{equation}
is the light intensity projected on the $i$-th node. In Eqs. 4 to 7, $\epsilon_1$, $\epsilon_2$, $\epsilon_3$, $\alpha$, $\beta$, $\gamma$, $\mu$, and $q$ are kinetic and time scale parameters, $I_0$ is the background light intensity, $K$ is the coupling strength, $L_{ij}$ is the Laplacian matrix of the corresponding coupling network, while $\xi_i (t)$ denotes white Gaussian noise.

We use the FHN model~\cite{IZH07} to study the influence of the relaxation character of the oscillators on the order of the synchronization transition. In a similar fashion to the ZBKE model, each oscillator is characterized by two dynamical variables: an ``activator'' ($u_i$) and an ``inhibitor'' ($v_i$) whose time evolution is given by
\begin{align}
	\dot{u}_i &= \epsilon \left( u_i - \frac{u_i^3}{3} - v_i + I_i \right), \\
	\dot{v}_i &= u_i + a,
\end{align}
where $a$ and $\epsilon$ represent dynamical parameters and the feedback
\begin{equation}
	I_i = K \sum_{j=1}^{N} L_{ij} \left[u_j + \xi_j(t) \right]
\end{equation}
acts additively on each node. The symbols $K$, $L_{ij}$, and $\xi_i(t)$ have the same meanings as in Eq. 7. All the parameter values are given in tables S1 and S2.

\textbf{Acknowledgments:} We acknowledge discussions with M. B\"{a}r, S. Yanchuk, and W. J. A. Martin. We thank U. K\"{u}nkel for support in the preparation of the experiments. \textbf{Funding:} J.F.T. and H.E. thank SFB 910 and GRK 1558. D.C. and J.F.T. thank DAAD RISE 2017, and D.C. thanks Trinity College, Cambridge for Trinity Summer Studentship Scheme 2017. \textbf{Author contributions:} D.C. and J.F.T. devised the study and did experimental and numerical work. D.C., J.F.T., E.A.M., and H.E. discussed the results, commented extensively on the manuscript at all stages of preparation, and jointly wrote the manuscript. \textbf{Competing interests:} The authors declare that they have no competing interests. \textbf{Data and materials availability:} All data needed to evaluate the conclusions in the paper are present in the paper and/or the Supplementary Materials. Additional data related to this paper may be requested from the authors.
\nocite{JAM01,MAR08,CAN10}
\bibliographystyle{apsrev4-2}
\bibliography{FirstSyncBib}

\end{document}